\documentclass[12pt]{article}

\usepackage{newtxtext,newtxmath}

\usepackage{graphicx}

\usepackage[letterpaper,margin=1in]{geometry}

\renewenvironment{abstract}
	{\quotation}
	{\endquotation}

\date{}

\makeatletter
\renewcommand{\fnum@figure}{\textbf{Figure \thefigure}}
\renewcommand{\fnum@table}{\textbf{Table \thetable}}
\makeatother

\usepackage{scicite}

\usepackage{url}

\makeatletter
\newcommand{\doublehat}[1]{%
\begingroup%
  \let\macc@kerna\z@%
  \let\macc@kernb\z@%
  \let\macc@nucleus\@empty%
  \hat{\mathchoice%
    {\raisebox{.3ex}{\vphantom{\ensuremath{\displaystyle #1}}}}%
    {\raisebox{.3ex}{\vphantom{\ensuremath{\textstyle #1}}}}%
    {\raisebox{.16ex}{\vphantom{\ensuremath{\scriptstyle #1}}}}%
    {\raisebox{.14ex}{\vphantom{\ensuremath{\scriptscriptstyle #1}}}}%
    \smash{\hat{#1}}}%
\endgroup%
}
\makeatother

\newcommand{\var}[1]{
    {\textbf{\texttt{#1}}}
}

\def\scititle{
	Quantum Magnetic $J$-Oscillators
}
\title{\bfseries \boldmath \scititle}

\author{
	Jingyan\,Xu$^{1,2,3}$,
	Raphael\,Kircher$^{1,2,3}$,
	Oleg\,Tretiak$^{1,2,3}$,
    Dmitry\,Budker$^{1,2,3,4}$,\and
    and Danila\,A.\,Barskiy$^{1,2,3\ast}$\and
	\small$^{1}$Institute of Physics, Johannes Gutenberg University of Mainz, 55099 Mainz, Germany.\and
	\small$^{2}$Helmholtz Institute Mainz, 55099 Mainz, Germany.\and
    \small$^{3}$GSI Helmholtzzentrum fur Schwerionenforschung, 64291 Darmstadt, Germany.\and
    \small$^{4}$Department of Physics, University of California, Berkeley, CA 94720, USA.\and
	\small$^\ast$Corresponding author. Email: dbarskiy@uni-mainz.de\and
}

\begin{document} 

\maketitle

\begin{abstract} \bfseries \boldmath
We introduce quantum $J$-oscillators that exploit intrinsic nuclear spin–spin (scalar $J$) couplings in molecules to produce phase-coherent oscillations. Operated in zero magnetic field and driven by a digital feedback, they operate from sub-hertz to a few tens of hertz frequencies. In a proof-of-principle experiment on [$^{15}$N]-acetonitrile,  the oscillator produced a 337\,$\mu$Hz linewidth over 3000\,s, more than two orders narrower than in conventional zero-field NMR. This may facilitate precision measurements of $J$-coupling constants and allows distinguishing mixtures of molecules whose zero-field NMR spectra would otherwise be hard to separate. In addition, the combination of strongly coupled spin systems and programmable feedback turns the $J$-oscillator into a compact tabletop (and, eventually, chip-scale) platform for exploring nonlinear spin dynamics, including chaos, dynamical phase transitions, and perhaps time-crystal behavior. By uniting high-resolution spectroscopy and controllable quantum dynamics in a single, magnet-free setup, $J$-oscillators open new opportunities for applications where ultraprecise frequency references or molecular fingerprints are required.

\end{abstract}

\noindent

\subsection*{Introduction}
Masers and lasers have revolutionized science and technology, finding applications in fields as diverse as telecommunications, medical diagnostics, astronomy, precision measurements, and fundamental physics \cite{gordon1955maser,prokhorov1958molecular ,schawlow1958infrared,kao1966dielectric,huang1991optical,thompson1987experiments,abbott2016observation}. Both technologies harness coherent electromagnetic radiation amplified through the process of stimulated emission\cite{einstein1917quantum}, a phenomenon that traditionally requires achieving population inversion between quantized energy states (i.e., a higher-energy state has to be more populated than a lower-energy state) \cite{svelto2010principles}.

Conventional masers have been realized across diverse physical systems, including atomic beams \cite{gordon1955maser, goldenberg1960atomic}, gases \cite{davidovits1966optically, chupp1994spin}, and solid-state systems\cite{bloembergen1956proposal,kraus2014room,breeze2018continuous}. These devices typically operate in the GHz frequency range, achieving the population inversion via various mechanisms. Recent advances extended these principles to the so-called ``rasers'' generating kHz-to-MHz frequencies using nuclear spins \cite{suefke2017hydrogen}. To create the required population inversion, such low-frequency rasers rely on hyperpolarization approaches such as spin-exchange optical pumping \cite{chupp1994spin}, dynamic nuclear polarization \cite{weber2019dnp},  photochemical polarization transfers \cite{buchachenko1996chemical}, and parahydrogen-based techniques \cite{suefke2017hydrogen, appelt2019laser, appelt2021sabre, lehmkuhl2022raser, nelson2025raser}. Additionally, their emission amplification is typically triggered by ``radiation damping'' \cite{bloembergen1954radiation}, stemming from inductive coupling of polarized nuclear spins with detection coils. However, these rasers exhibit intrinsic limitations as they operate on Zeeman-split levels. Since the frequencies of these levels depend on the applied (bias) magnetic field, they are susceptible to magnetic field drifts, limiting their long-term stability and reproducibility.

In this work, we introduce zero-field quantum oscillators 
operating at frequencies from near-DC to tens of hertz, overcoming limitations of conventional rasers that rely on Zeeman-split levels. Unlike those, our oscillators function without a bias field, exploiting intrinsic nuclear spin-spin scalar interactions ($J$-couplings) within molecules \cite{blanchard2013high, barskiy2025zero}. Importantly, they utilize $\Delta m = 0$ transitions \cite{barskiy2025zero}, with the quantization axis along the measurement axis (Fig.\,\ref{fig:schematics}A,B). Because these transition frequencies depend primarily on molecular $J$-coupling constants \cite{barskiy2025zero}, the zero-field quantum oscillators achieve significantly improved frequency stability: coherent operation of up to 1\,h is demonstrated (Fig.\,\ref{fig:schematics}D). The magnetic oscillation occurs along the same axis as the feedback axis due to $\Delta m = 0$ transitions \cite{barskiy2025zero}, unlike the precessing magnetization with $\Delta m = \pm 1$ of conventional rasers. Such an oscillator requires only a population imbalance—not a strict population inversion—achieved \textit{in situ} by bubbling parahydrogen (\textit{para}-H$_2$) into a liquid containing the activated  SABRE catalyst (SABRE = signal amplification by reversible exchange), see Fig.\,\ref{fig:schematics}C \cite{adams2009reversible, kircher2025benchtop}. The catalyst enables spontaneous polarization transfer \textit{in situ} at zero field, creating $J$-transition population imbalances in target molecules \cite{theis2012zero, xu2025zero}. Details of the \textit{para}-H$_2$ gas handling are given in the Supporting Information (SI, Material and Methods).

\subsection*{External programmable feedback loop}
A central challenge in implementing zero-field quantum oscillators arises from the absence of radiation damping. In ultralow-field raser experiments—also lacking radiation damping—researchers have relied on external feedback loops \cite{sato2018development}. Typically, these feedback loops detect precessing magnetization using optically pumped magnetometers (OPMs) \cite{budker2007optical}, and subsequently feed signals back into the sample through coils orthogonal to the measurement axis \cite{jiang2021floquet}. However, such feedback scheme cannot be directly adopted to zero-field $J$-oscillators. This is because coherent amplification of the $\Delta m = 0$ transitions requires that the feedback magnetic field is applied along the same measurement axis ($y$-axis, \cite{barskiy2025zero}). 

To resolve this, we developed an external feedback loop, implemented via software control without need for specialized hardware modifications as schematically depicted in Fig.\,\ref{fig:schematics}A. In our implementation, an OPM detects the $y$- cartesian component of the magnetic field generated by the sample ($B_{\mathrm{OPM}}$). This signal is processed digitally, allowing precise control of both a tunable external feedback delay ($\tau$) and feedback gain ($G_{\mathrm{ext}}$). The processed feedback signal is reapplied to the sample ($B_{\mathrm{ext}}$, along the $y$-axis) using a piercing solenoid \cite{yashchuk2004hyperpolarized}. The piercing solenoid configuration inherently avoids feedback field leakage into the OPM sensor ensuring that measured signals are exclusively generated by the sample.

\subsection*{Stability of quantum oscillations}
The quantum oscillator was initially tested on a model system consisting of 5\,\% [$^{15}$N]-acetonitrile ([$^{15}$N]-ACN) dissolved in acetonitrile (ACN) solvent \cite{xu2025zero, kircher2025benchtop}. The feedback configuration enables spontaneous emergence of the quantum oscillator, as demonstrated in Fig.\,\ref{fig:schematics}D with a feedback delay of $\tau=222$\,ms and gain $G_{\mathrm{ext}}=+20$, requiring no pulse excitation. Likely, the electronic noise from the feedback loop triggers initial transitions. The signal from these spontaneous transitions is amplified by the feedback loop and the phase is tuned for positive reinforcement. The resulting amplified field ($B_{\rm ext}$) drives the transitions, pushing the population imbalance away from the SABRE-pumped hyperpolarized steady state \cite{hovener2013hyperpolarized}. With high enough external feedback gain, this SABRE-pumped population imbalance can be temporarily inverted (see Fig.\,\ref{fig:overshooting}). Deviations from the hyperpolarized steady state manifest experimentally as transient bursts. Such bursts represent the so called ``overshooting,'' where the feedback-amplified transition intensities rise beyond  sustainable levels \cite{jiang2021floquet, appelt2019laser}. The SABRE pumping and nuclear spin relaxation processes oppose the feedback amplification, dampening coherences and partially restoring the hyperpolarized steady state. Over successive faded bursts, the system stabilizes into what we call the ``dynamic steady state'' under the feedback, in which the combined effects of SABRE-pumping (replenishing population imbalances), the feedback-amplified transitions (driven by $B_{\mathrm{ext}}$), and relaxation come into a balance.

The quantum oscillator operating on the molecular $J$-transition was continuously recorded for over 3000\,s. Fourier transformation (FT) (Fig.\,\ref{fig:schematics}E) revealed a sharp, delta-function-like peak with a full-width-at-half-maximum (FWHM) of 337\,\(\mu\)Hz (Fig.\,\ref{fig:schematics}E, inset). This linewidth 
is approximately the inverse of the measurement time, indicating minimal frequency drift of the experimentally observed quantum oscillation. 
Consequently, the linewidth of such stable quantum oscillations scales inversely with measurement duration, suggesting prolonged measurements may yield even narrower FWHM.  In contrast, the linewidth obtained from conventional zero-field NMR spectra does not improve with increased measurement time, as it is fundamentally limited by nuclear spin relaxation. For instance, the [$^{15}$N]-ACN has a FWHM of 37\,mHz for the same transition (Fig.\,\ref{fig:schematics}H). Similarly to high-field rasers \cite{nelson2025raser},  the reduced linewidth achieved with the $J$-oscillators facilitates resolving overlapping resonance lines.



While both $J$-oscillators and conventional zero-field $J$-spectra demonstrate the same scaling behavior of SNR with total acquisition time ($\text{SNR}\propto \sqrt{T}$, see Fig.\,\ref{fig:scaling}), the error in determining transition frequency using $J$-oscillator approach decreases more rapidly with the acquisition time due to the above-mentioned linewidth narrowing. Therefore, standard deviation in measuring precision in frequency space using $J$-oscillators approaches Cramér-Rao lower bound \cite{fleischer2023approaching}.


\subsection*{On-demand spectral editing}
For a negative external feedback gain ($G_{\mathrm{ext}} = -20$, Fig.\,\ref{fig:delay and gain}C), the 1-$J$ quantum oscillator is sustained for feedback delays ranging from 60 to 275\,ms (minimal delay in our system is 60 ms). Under the same negative feedback gain conditions, the 2-$J$ quantum oscillator emerges only within a narrower delay range of 60–140\,ms.  For positive external feedback gain ($G_{\mathrm{ext}} = 20$, Fig.\,\ref{fig:delay and gain}D), the 1-$J$ oscillator emerges spontaneously at delays ranging from 325 to 400\,ms (400\,ms being the maximum sampled delay in our system). Meanwhile, the 2-$J$ oscillator is sustained under delays spanning from 160\,ms to 290\,ms. Thus, the emergence of 1-$J$ and 2-$J$ quantum oscillators exhibit different dependencies on the externally applied feedback delays. 
The fact of digital (as opposed to analog) feedback can open more possibilities in the future to control the phase lag/advance and external gain for different peaks separately.

This behavior arises due to the frequency-dependent phase shifts introduced by the delays. To gain more insight, the delay ($\tau$) is converted to the corresponding feedback phase lag ($\varphi$) at the operating $J$-transition frequency $f$, calculated as $\varphi = 2 \pi f \tau$ as presented in Table\,\ref{tab:phase_lags}. 
Combining the results from both plots, it was found that the 1-$J$ quantum oscillator is sustained across the range of phase lags from around 3.4 to 6.1\,radians. Similarly, the 2-$J$ oscillator remains active within a comparable phase lag interval, approximately between approx. 3.4 and 6.1~radians. 
These phase intervals are symmetric around $3\pi / 2$, spanning roughly $ \pm 0.86 \pi / 2$ around this central value.

The dependence of the $J$-oscillator operation on phase can be understood by decomposing the feedback field mathematically into two components relative to the field generated by the sample. The first component aligns with the signal (“in-phase,” phase lag 0) or is opposite to it (“anti-phase,” phase lag $\pi$). The second component, the “quadrature” component, is shifted by a quarter cycle (phase lag $\pm \pi / 2$) relative to the signal. Only this quadrature component effectively contributes to amplifying the selected $J$-transitions. To quantify this amplifying contribution, we define an “effective external feedback gain,” represented as $G_{\mathrm{ext}}\sin\varphi$ \cite{chacko2024multimode}. For sustained quantum oscillations to emerge, the absolute value of this effective gain must exceed a specific threshold (discussed below). Moreover, the sign of the effective gain must match the direction of the hyperpolarized population imbalances: in the current case, a negative effective gain is required, corresponding to a population inversion between the selected $J$-transitions. Consequently, the dependence of quantum oscillator emergence on the externally imposed phase lag provides a convenient method to selectively amplify individual $J$-transitions at specific frequencies, demonstrating the capability for on-demand spectral editing (see Fig.\,\ref{fig:diverse molecules}).

\subsection*{Threshold dynamics}
Figures\,\ref{fig:delay and gain}E and\,\ref{fig:delay and gain}F show the dependence of the steady-state and the initial burst amplitudes on the external feedback gain ($G_\mathrm{ext}$) for quantum oscillators operating selectively at the 1-$J$ (Fig.\,\ref{fig:delay and gain}E, $\tau\!=\!160$\,ms) and 2-$J$ (Fig.\,\ref{fig:delay and gain}F, $\tau\!=\!222$\,ms) transitions, respectively. These delays ensure that the feedback corresponds to a phase lag near $3\pi/2$ for each peak, resulting in a maximal contribution of the quadrature component ($|\sin\varphi|=1$).  In both plots, the steady-state amplitudes initially rise with increasing the feedback gain but decline at higher gains, mirroring the conventional maser/laser dynamics where output peaks at an optimized resonator quality factor \cite{siegman1986lasers}. The fact that higher gains result in lower SS amplitudes is also indicated by the minima observed in Fig.\,\ref{fig:delay and gain}C (1-$J$ at 160\,ms) and \ref{fig:delay and gain}D (2-$J$ at 220\,ms). These minima correspond to conditions with the feedback phase lag of approx. $3\pi/2$ where the effective feedback gain has the biggest amplitude. At higher gains, the maximal amplitude of the first burst plateaus as the hyperpolarization-driven population imbalance is converted into coherences.

The observed $J$-oscillations in Fig.\,\ref{fig:delay and gain}E-F are affected by two contributions: the  passive magnetic intrinsic gain of the system $G_{\mathrm{int}}$ and the actively applied external feedback gain $G_{\mathrm{ext}}$. The intrinsic gain $G_{\mathrm{int}}$, which is analogous to the conventional gain definition in laser physics \cite{svelto2010principles}, quantifies the  ability of the system to amplify magnetic fields. Specifically, the intrinsic gain $ G_{\mathrm{int}} $ is defined as the ratio between the amplitude of the magnetic field produced by the sample (as measured by the optical magnetometer, OPM) and the amplitude of the externally applied AC field that acts on the sample (see SI for additional details).
For sustained quantum oscillations to emerge spontaneously, the total gain from combined internal and external contributions must exceed unity:
\begin{equation}
|G_{\mathrm{ext}} \cdot G_{\mathrm{int}}| > 1\,.\,\label{eq:Gth}
\end{equation}
For the cases shown in Figs.\,\ref{fig:delay and gain}E and \,\ref{fig:delay and gain}F, the threshold external gains $G^{\rm th}_{\rm ext}$ required to initiate sustained oscillations, extracted from the numerical simulation illustrated in figures, are approximately 7.2 for the 1-$J$ transition and 5.8 for the 2-$J$ transition, respectively, as highlighted by the shaded areas in the respective figures. Additionally, simulations of the exact same modeled spin systems yield intrinsic gain values $G_{\mathrm{int}}$ of approximately 0.138 and 0.172 for the 1-$J$ and 2-$J$ transitions, respectively (see Fig.\,\ref{fig:Gin} and SI for detailed numerical procedures). Overall, these extracted values meet the theoretical threshold condition defined by Eq.\,\ref{eq:Gth}.


By tunning the external feedback gain, it is possible to achieve quantum oscillators under challenging conditions such as low molar polarization (product of molecular concentration and nuclear polarization) or rapid relaxation, where intrinsic gain alone would be insufficient to initiate the emergence of the oscillator\cite{suefke2017hydrogen}. Increasing the external feedback gain ($G_\mathrm{ext}$) effectively lowers the required polarization threshold, allowing the observation of the quantum oscillator behavior for various molecules, see below. Conversely, when the external feedback gain is limited, only the $J$-transitions with sufficiently large intrinsic gain can surpass the threshold condition. Exploiting this property enables selective excitation of individual resonance lines within spectra containing densely overlapping peaks (see Fig.\,\ref{fig:diverse molecules}).


\subsection*{Generality across various molecules}
Figure\,\ref{fig:diverse molecules} presents spectra of $J$-oscillators alongside their corresponding zero-field NMR spectra for various illustrative chemical systems. All $J$-oscillator spectra were acquired over 10\,min (except for Figure\,\ref{fig:diverse molecules}F, which was acquired over approx. 20\,s). The corresponding FT spectra were normalized using division by the total number of data points (Convention I, see SI), allowing for the direct comparison between quantum oscillator-based and zero-field NMR spectra, despite differences in acquisition times. The preparation of the samples is discussed in the SI (Material and methods section).

Figure 3A shows $J$-oscillators generated from naturally abundant [$^{15}$N]-acetonitrile ([$^{15}$N]-ACN, 0.36\,\%). Despite roughly ten-fold isotopic dilution compared to the previously discussed samples (5\,\%), we successfully achieved sustained oscillations by compensating the reduced intrinsic gains with increased external digital feedback gain.  Additionally, isotopically labeled ACN molecules with 1\,\% abundance of $^{15}$N were tested: specifically, [1-$^{13}$C,$^{15}$N]-ACN (Fig.\,\ref{fig:diverse molecules}B) and [2-$^{13}$C,$^{15}$N]-ACN (Fig.\,\ref{fig:diverse molecules}C). We also investigated a more complex nitrile (Fig.\,\ref{fig:diverse molecules}I), [U-$^{13}$C,$^{15}$N]-butyronitrile, whose conventional zero-field spectrum (blue trace) is significantly complicated by a complex interplay of spin-spin couplings. By carefully tuning the feedback parameters, stable $J$-oscillations were achieved on individual transitions, demonstrating the potential for systematic on-demand spectral editing. 

Notably, a ``near-DC'' signal was detected for [U-$^{13}$C, $^{15}$N]-butyronitrile under certain conditions, i.e., a static magnetization under feedback that persisted after emergence for more than ten minutes. 
These DC (or near-DC) frequency signals are unresolvable by the conventional zero-field NMR because relaxation rates for these transitions exceed their coherent oscillation frequencies while the extended measurement window enabled us to observe them. Whether it is a low frequency oscillations or a some sort of nuclear paramagnetic phenomenon warrants additional investigations.

Apart from nitriles, heterocyclic molecules such as [$^{15}$N]-pyridine (Fig.\,\ref{fig:diverse molecules}D), 4-amino[$^{15}$N]-pyridine(Fig.\,\ref{fig:diverse molecules}E), [$^{15}$N$_3$]-metronidazole (Fig.\,\ref{fig:diverse molecules}F), and [$^{15}$N$_2$]-imidazole  (Fig.\,\ref{fig:diverse molecules}G) were explored. These molecules were available in the lab and can be readily polarized with SABRE. While conventional zero-field $J$-spectra of these molecules suffer from the severe spectral overlap and broad linewidths, the quantum oscillators instead yield narrow, mHz-scale resonances that selectively address individual transitions. The demonstrations of quantum oscillators operating on [$^{15}$N]-pyridine under a higher feedback gain ($G_{\mathrm{ext}}=-3000$) 
is given in  Fig.\,\ref{fig:pyridine_oscillator}. In this strong feedback regime, oscillations corresponding to $J$-transitions at approximately 1\,Hz, 4\,Hz, and 10\,Hz become clearly observable. Furthermore, the dynamics of the $J$-transition at around 15\,Hz under the elevated feedback gain are presented, matching predictions from simplified numerical simulations (Fig.\,\ref{fig:2spin_simu}). Likewise, the quantum oscillators emitting at ultralow frequencies (below 0.5\,Hz) were observed for [$^{15}$N$_3$]-metronidazole and 4-amino-[$^{15}$N]-pyridine (see Fig.\,\ref{fig:heterocyclic_ultralow}).

Finally, we demonstrate the $J$-oscillator using [1-$^{13}$C]-pyruvate (Fig.\,\ref{fig:diverse molecules}H). Due to the poor hyperpolarization at elevated temperature conditions (heat up by the OPM in our measurement apparatus), together with the broad Hz-scale FWHM, the conventional zero-field NMR spectrum of this molecule (blue trace) exhibited poor SNR ($\sim$4). In order to achieve the quantum oscillator for this molecule, an external feedback gain of $G_\mathrm{ext}= 50,\!000$ was applied. The resulting quantum oscillator (orange trace) showed significantly improved SNR and a narrower linewidth. Notably, a frequency shift of both peaks was detected, because the feedback loop coupled the two transitions and ``drew'' the peaks toward each other \cite{chacko2024multimode}.

\subsection*{Application}

To illustrate the analytical power of the quantum $J$-oscillator, we investigated a series of mixtures that all contain 50~mM pyridine, 50~mM 4-aminopyridine and 3~mM of the Ir-based catalyst, but differed in the $^{14}$N/$^{15}$N isotopic ratio of the two substrates. Conventional zero-field NMR spectra of these samples show that both molecules have their main peaks overlapping at around 15~Hz (Fig.\,\ref{fig:mix_fid}). In contrast, the $J$-oscillator approach yields well-resolved peaks for each molecule. Figure\,\ref{fig:mixture} shows the resulting spectra, all obtained using an identical feedback delay ($\tau = 115$ ms) but varying feedback gain.

For the mixture in which both substrates were 100~\% $^{15}$N-labeled, two narrow resonances with about identical intensities emerge at moderate feedback gains. Comparison with the entire set of data reveals that the lower-frequency peak originates from 4-amino-[$^{15}$N]-pyridine whereas the higher-frequency line stems from [$^{15}$N]-pyridine. With high enough feedback gain, the $J$-oscillations from molecules that are 50~\% $^{15}$N-labelled can also be obvserved (Fig.\,\ref{fig:mix_SI}). In the sample that contained 100~\% [$^{15}$N]-pyridine and 50~\% 4-amino-[$^{15}$N]-pyridine, the right peak is about twice as strong as the left one, giving the expected 2:1 ratio that reflects upon its isotopic composition. The opposite 1:2 intensity ratio is observed when the enrichment levels are swapped.

The experiments also show several interesting dynamical effects. First, whenever the feedback gain is tuned such that only one molecule oscillates, its resonance drifts to higher frequencies as the gain is increased. Second, when both oscillators are excited, the two resonances from the two molecules push each other apart and the frequency separation increases with gain. These shifts do not reflect changes in the intrinsic $J$-couplings; they arise because the feedback field loop couples the spin dynamic of the two molecules. A detailed quantitative treatment is beyond the scope of the present work and will be reported elsewhere. Note the shifts are reproducible and can be calibrated by a careful tuning of the feedback parameters so that the intrinsic $J$-couplings can be extracted. Consequently, this method opens avenues toward analyzing complex mixtures containing structurally similar molecules whose resonance signals typically overlap substantially in conventional zero-field NMR experiments.

As the applied feedback gain is further increased, a rich variety of nonlinear dynamics—including the emergence of frequency combs and chaos—arises from the imposed feedback that couples different $J$-transitions. These phenomena highlight the unique potential of zero-field $J$-oscillators as convenient platforms for systematic exploration of nonlinear spin dynamics.

A key advantage of the zero-field $J$-oscillator (compared to Zeeman Rb-Xe co-magnetometers \cite{jiang2021floquet} or high-field rasers \cite{suefke2017hydrogen}) is its ability to operate as an engineered multi-mode system. Apart from requiring mixtures of structurally similar molecules, this can be achieved by applying a static bias field, which splits a single $J$-transition into multiple transitions and allows continuous tuning of their frequency separations simply by adjusting the field amplitude (to be published elsewhere). Such tunability can enable the systematic exploration of phase transitions in the system’s dynamics--from mode collapse, to frequency combs, to chaos--as the bias field is varied. In contrast, high-field rasers lack this flexibility, as their mode spacing cannot be easily tuned (e.g. intrinsic $J$-couplings are fixed).

Digital processing also enables the inclusion of advanced algorithms for narrow-band filtering, enhancing spectral selectivity and suppressing experimental imperfections such as OPM temperature drift. Moreover, it allows for the implementation of more sophisticated feedback schemes—including those based on various derivatives or polynomial functions of the detected signal. These capabilities facilitate controlled studies of complex phenomena, ranging from chaos to the realization of molecular spin-based time crystals \cite{zaletel2023colloquium}.


\subsection*{Conclusion}
In analogy to conventional lasers, the coherent signal generated by $J$-oscillators presented in this work significantly boosts resolution in zero-field NMR spectroscopy by achieving exceptionally narrow linewidths. Furthermore, due to intrinsic insensitivity of $J$-coupling constants to magnetic field drifts, these $J$-oscillators exhibit superior long-term frequency stability in comparison to conventional NMR rasers. 
The selective resonance amplification capability (on-demand spectral editing) provided by $J$-oscillators enables individual transitions to be excited and resolved, even when spectra are composed of densely overlapped resonances. 
The enhanced resolution of overlapping spectral features demonstrates the power of $J$-oscillator approach to distinguish structurally similar molecules in mixtures. 
By taking advantage of the ability of applying the external feedback gain to lower polarization threshold requirements, the $J$-oscillations were demonstrated for a diverse range of molecules, even in samples with natural isotopic abundance. 
Beyond the analysis of mixtures, the ability to manipulate the digital programmable feedback operating on coupled spin systems at zero field opens opportunities for exploring nonlinear spin dynamics, including phenomena such as chaos and time crystals. These capabilities highlight the broader potential of quantum $J$-oscillators for advanced analytical applications, probing fundamental physics, and for studies of nonlinear spin dynamics.

\begin{figure} 
	\centering
	\includegraphics[width=15cm]{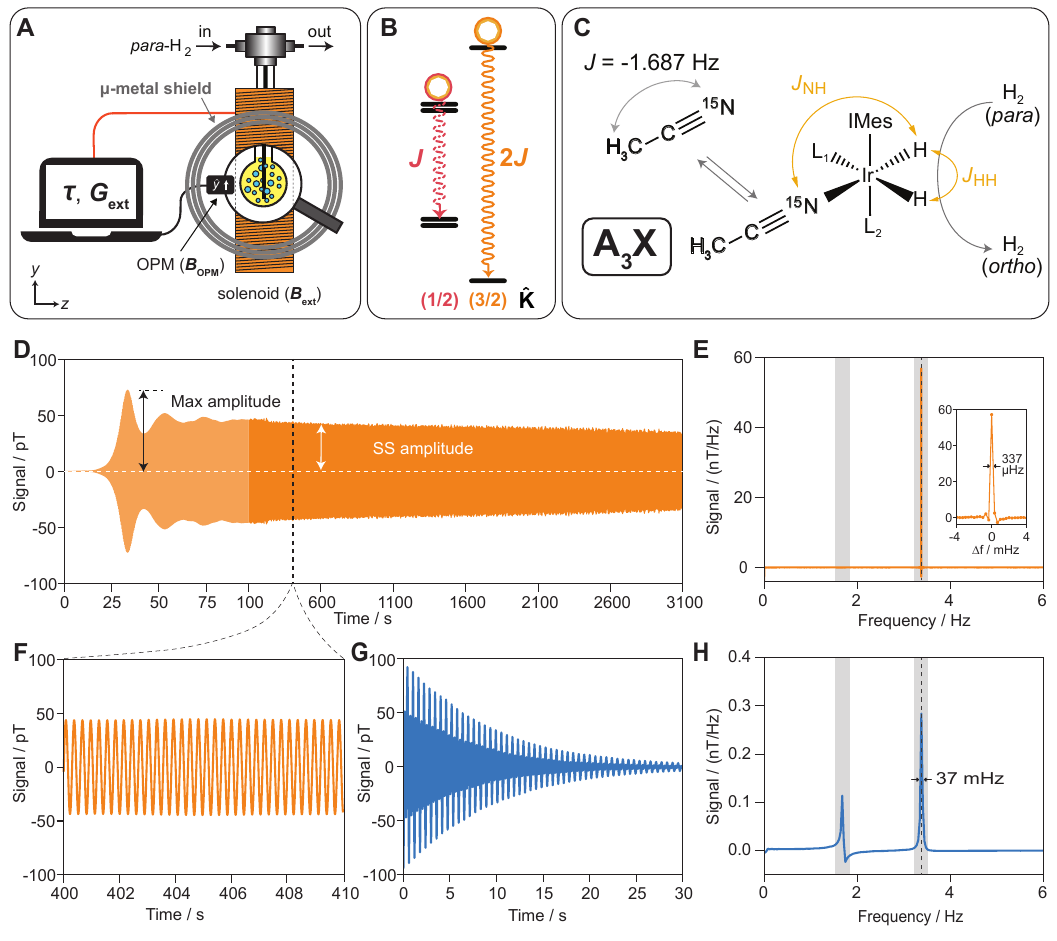}     
	\caption{\textbf{The concept of a zero-field $J$-oscillator.} (A) Schematic representation of the real-time programmable feedback loop used for observing $J$-oscillators; the signal from hyperpolarized molecules is detected by an optically pumped magnetometer (OPM), processed digitally with defined feedback delay ($\tau$) and gain ($G_{\mathrm{ext}}$), and reapplied to the molecules via a piercing solenoid.
    (B)~Energy-level diagram of the $J$-coupling transitions of [$^{15}$N]-acetonitrile ([$^{15}$N]-ACN) at zero field.
    (C)~[$^{15}$N]-ACN is hyperpolarized \textit{in situ} at zero field via SABRE; the Ir-based catalyst facilitates spontaneous spin-order transfer from parahydrogen (\textit{para}-H$_2$) to the population imbalances in the target molecules. (D)~Experimentally recorded quantum oscillator time-domain signal; observation of the spontaneous emergence ($<100$\,s) and subsequent steady-state (SS) coherent oscillation is obtained with $\tau=222$\,ms and $G_{\mathrm{ext}}=+20$. (E) Real part of the Fourier transform of the SS oscillation (100–3100\,s); the inset shows a $\delta$-function-like peak at the 2-$J$ frequency with a measurement-time-limited FWHM of 337\,$\mu$Hz ($\Delta f$ is referenced to 3.374\,Hz). (G-H) conventional time-domain and frequency-domain zero-field NMR signals from the same sample.  }
	\label{fig:schematics} 
\end{figure}

\begin{figure} 
	\centering
	\includegraphics[width=8.64cm]{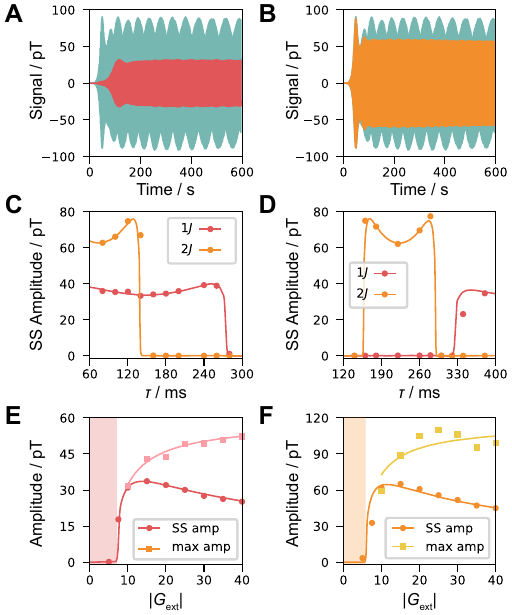}     
	\caption{\textbf{Feedback-delay-dependent spectral selectivity and threshold dynamics of zero-field $J$-oscillators.} (A)-(B) Oscillator on [$^{15}$N]-acetonitrile (blue) obtained with parameters $\tau=100$\,ms and $G_{\mathrm{ext}}=-20$ demonstrates coherent generation from both 1-$J$ and 2-$J$ transitions; Fourier filtering after processing isolates 1-$J$ oscillation (red) in (A) and 2-$J$ oscillation (orange) in (B). Panels (C) and (D) 
    show the steady-state (SS) amplitudes of 1-$J$ and 2‑$J$ oscillation versus delay ($\tau$), with a fixed feedback gain of $G_{\mathrm{ext}}=-20$ (C) and $G_{\mathrm{ext}}=+20$ (D), respectively. Panels (E) and (F) show the SS and maximal amplitude dependence on the feedback gain $G_{\mathrm{ext}}$ for  oscillators operating exclusively on 1-$J$ (negative feedback, $\tau=160$\,ms) and 2-$J$ (positive feedback, $\tau=222$\,ms) transitions, respectively.  Shaded areas in (E) ($G_{\mathrm{ext}}<7.2$) and (F) ($G_{\mathrm{ext}}<5.8$) indicate regions where the feedback gains fall below the threshold needed for the spontaneous oscillator emergence.
    For signals not reaching steady state within 600\,s, the amplitude of the last 10 s was used. Solid lines in (C-F) are actual numerical simulations with the algorithm described in the SI. }
	\label{fig:delay and gain} 
\end{figure}

\begin{figure} 
	\centering
	\includegraphics[width=15.4cm]{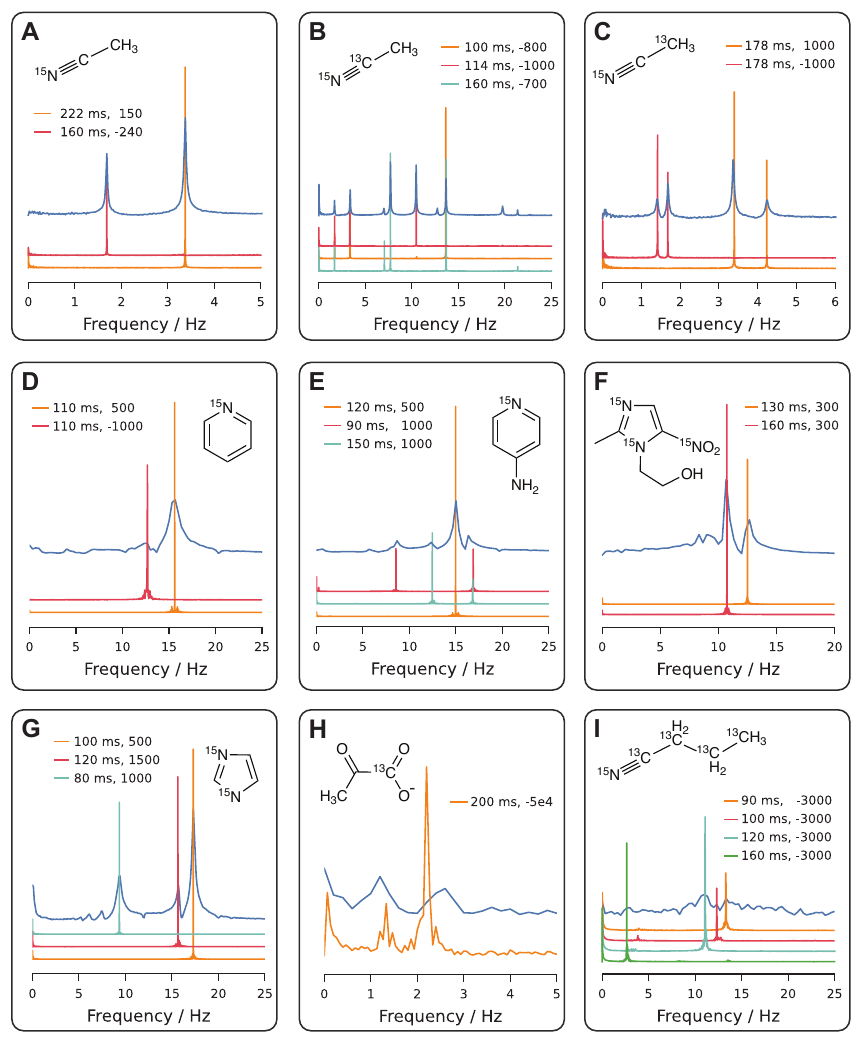}     
	\caption{\textbf{Quantum $J$-oscillators realized on various molecules.} Comparison between conventional zero-field NMR spectra (blue traces) and quantum oscillator signals from exemplary chemical systems: acetonitrile (ACN) solvent with various labeling percentages: (A) natural abundance (0.36\,\% and 99.64\,\%); (B) 1$\%$ [2-$^{13}$C,$^{15}$N]-ACN in [$^{14}$N]-ACN; (C) 1$\%$ [1-$^{13}$C,$^{15}$N]-ACN in [$^{14}$N]-ACN; (D) [$^{15}$N]-pyridine; (E) 4-amino[$^{15}$N]-pyridine; (F) [$^{15}$N$_3$]-metronidazole; (G) [$^{15}$N$_2$]-imidazole; (H) [1-$^{13}$C]-pyruvate; (I) [U-$^{13}$C, $^{15}$N]-butyronitrile.}
	\label{fig:diverse molecules} 
\end{figure}

\begin{figure} 
	\centering
	\includegraphics[width=8.4cm]{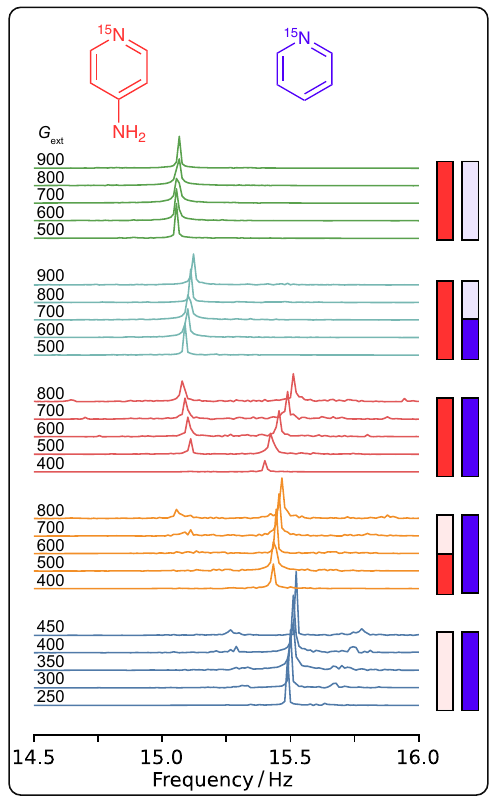}     
	\caption{\textbf{Quantum $J$-oscillators operated on pyridine/4-aminopyridine mixture with different $^{14}$N/$^{15}$N isotopic compositions.} Each sample contains 50~mM pyridine, 50~mM 4-aminopyridine, and 3~mM of the Ir-based catalyst in methanol, with only the $^{15}$N isotopic enrichment of the two substrates varied. For each sample, the amount of substrate is indicated by a pair of vertical bars: the left red bar for 4-aminopyridine and the right blue bar for pyridine. The shading of each bar denotes the degree of $^{15}$N enrichment: solid color corresponds to 100~\% $^{15}$N, fully transparent corresponds to natural-abundance $^{14}$N, and a half solid/half transparent bar corresponds to 50~\% $^{15}$N enrichment. The traces measured on the same sample are plotted with the same color, with the feedback delay fixed at $\tau=115\,\mathrm{ms}$ and the applied external feedback gain annotated to the left of each trace. The Fourier spectra are obtained by analyzing the final 90~s of the recorded 180~s of $J$-oscillation data in order to report on the steady state and avoid the burst/transient. For reference, the conventional zero-field NMR spectrum of each sample is provided in Fig.\,\ref{fig:mix_fid}. }
	\label{fig:mixture} 
\end{figure}

\newpage


\clearpage 

%
\bibliography{quantum_oscillator/main.bib} 
\bibliographystyle{sciencemag}

%
%
%
%
%
%


\section*{Acknowledgments}
\paragraph*{Funding:}
The work is supported by the Alexander von Humboldt Foundation in the framework of the Sofja Kovalevskaja award and by DFG/ANR grant BU 3035/24-1.
\paragraph*{Author contributions:}
Conceptualization: JX, RK, OT, DB, DAB; Investigation: JX, RK, DAB;
Funding acquisition: DB, DAB;
Project administration: DAB;
Writing – original draft: JX;
Writing – review $\&$ editing: JX, RK, OT, DB, DAB.
\subsection*{Supplementary materials}
Materials and Methods\\
Figs. S1 to S9\\
Tables S1 to S2\\
References (27, 29, \textit{38-\arabic{enumiv}})\\ 


\newpage


\renewcommand{\thefigure}{S\arabic{figure}}
\renewcommand{\thetable}{S\arabic{table}}
\renewcommand{\theequation}{S\arabic{equation}}
\renewcommand{\thepage}{S\arabic{page}}
\setcounter{figure}{0}
\setcounter{table}{0}
\setcounter{equation}{0}
\setcounter{page}{1} 


\begin{center}
\section*{Supplementary Materials for\\ \scititle}

    Jingyan\,Xu,
	Raphael\,Kircher,
	Oleg\,Tretiak,
    Dmitry\,Budker, \and 
    Danila\,A.\,Barskiy$^{\ast}$\and
    \\ 
\small$^\ast$Corresponding author. Email: dbarskiy@uni-mainz.de
\end{center}



\newpage


\subsection*{Materials and methods}
\subsubsection*{Parahydrogen handling}
The ZULF setup is presented in a simplified scheme in Figure\,\ref{fig:schematics}A. The \textit{p}H$_2$ gas-handling setup was presented in detail in previous work of our group \cite{SHEBERSTOV2025100194,kircher2025benchtop}. Controlled supply of enriched \textit{p}H$_2$ into the NMR tube containing the liquid sample was enabled by several pneumatic valves and a back-pressure regulator from Swagelok. For all experiments, hydrogen gas was first passed over a packed bed of an iron oxide catalyst, FeO(OH), at a temperature of 30\,K to selectively convert the commercially available di-hydrogen gas mixture into \textit{p}H$_2$ (approx. 97\% enrichment).
After conversion, the gas in our piping system (mainly made of polyetheretherketone) is warmed up to approx. 21\,°C at a pressure of 10\,bar before it is introduced into the NMR tube. Two Teflon capillaries were inserted into the NMR tube, one leading to the bottom and used for inflow of \textit{p}H$_2$ gas and the second capillary only at the top for the gas outlet. The design of NMR tubes is presented in the literature \cite{SHEBERSTOV2025100194, xu2025zero}. Flow of hydrogen gas into the NMR tube was controlled with a mass flow controller from Sierra Instruments (20\,scc min$^{-1}$). Quantum oscillators were measured with continuous parahydrogen bubbling. Zero-field NMR spectra were collected by pulsed bubbling of parahydrogen in short, controlled cycles, where each cycle involved 2.6\,s of bubbling followed by a 0.5-second pause, repeated five times in total. The so-called ``ZF-ZF'' \cite{xu2025zero} magnetic field sequences were adopted for collecting the conventional zero-field NMR spectra. The pressure in the NMR tube was controlled using a back-pressure regulator that was set to 7\,bar. Control of all timings, pneumatic valves, and communication with hardware required to set up ZULF detection were realized by TTL pulses programmed by Python.

\subsubsection*{External feedback loop apparatus}
The hardware components of the feedback loop system are as follows. The signal produced by the sample was measured using a commercially available optically pumped magnetometer (model QuSpin Gen\,2) with a sensitivity of around 15\,fT/$\sqrt{\mathrm{Hz}}$ in a 3-100\,Hz band. The analog output from the OPM was digitized with a National Instruments (NI) analog input card (model NI 9239) at a sampling rate of~2000 Hz. This digitized signal was transmitted to the computer via the NI Measurement $\&$ Automation Explorer interface. A Python-based real-time processing algorithm running on the computer set the delay and gain parameters to the acquired signal. The processed signal was then converted back to an analog form using an analog output (model NI 9263) at a sampling rate of 2000 Hz as well and delivered to a piercing solenoid, which generated the feedback magnetic field.

\subsubsection*{Sample preparation}
Samples were prepared in a nitrogen-atmosphere and transferred into a spherical NMR tube (outer diameter\,=\,10.5\,mm, inner diameter\,=\,8.5\,mm) \cite{xu2025zero} that can be adapted to the \textit{p}H$_2$ gas-handling setup. Samples were prepared of a common Ir-precursor [Ir(IMes)(COD)Cl] dissolved in acetonitrile or methanol, c.f. detailed sample composition is provided in Table\,\ref{tab:sample_compositions}. The following substrates and solvents were purchased from Sigma Aldrich and used without further purification: (A) acetonitrile (0.36\,\% [$^{15}$N]-ACN and 99.64\,\% [$^{14}$N]-ACN), (B) 1\,$\%$ [2-$^{13}$C,$^{15}$N]-ACN in 99\,\% [$^{14}$N]-ACN, and (C) 1$\%$ [1-$^{13}$C,$^{15}$N]-ACN in 99\,\% [$^{14}$N]-ACN; (D)[$^{15}$N]-pyridine, (E) 4-amino[$^{15}$N]-pyridine, (F)[$^{15}$N$_3$]-metronidazole, (G) [$^{15}$N$_2$]-imidazole, (H)[1-$^{13}$C]-pyruvate, and (I) [U-$^{13}$C, $^{15}$N]-butyronitrile, c.f. Figure\,\ref{fig:diverse molecules}. An additional co-substrate \cite{kircher2025benchtop, iali2019hyperpolarising, mewis2015strategies} was introduced to stabilize hyperpolarization transfer efficiency of samples (A,B,C,H, and I), which are listed in Table\,\ref{tab:sample_compositions}.

\subsubsection*{Programmable feedback algorithm}
Key parameters:
\begin{itemize}
    \item \var{READ}: Array storing voltage samples read from NI 9239.
    \item \var{WRITE}: Array storing voltage samples to be written to NI 9263.
    \item \var{Nc}: Number of samples per chunk (size of READ and WRITE).
    \item \var{DELAY} Feedback delay in the number of samples (set to the value of $\tau \times f_s$, where $f_s$ is the sampling rate in Hz minus a hardware delay).
    \item \var{Buffer}: The cyclic buffer array used for storing previously read data.
    \item \var{Nb}: Total size of the cyclic Buffer.
    \item \var{Gext}: The external feedback gain. To ensure the feedback operates with the correct gain, an appropriate resistor is connected in series between the output of the NI 9263 and the piercing solenoid.
    \item \var{i}: Index pointer for the current reading or writing position in the Buffer.
\end{itemize}

\paragraph{Initialization:}
\begin{itemize}
    \item Initialize the entire \var{Buffer} with zeros:\\
    \var{Buffer[:]=0}
    \item Initialize the buffer index pointer:\\
    \var{i=0}
\end{itemize}

\paragraph{Continuous operation loop:}
\begin{itemize}
    \item  Acquire Data: \\
    Read \var{Nc} voltage samples from the NI 9239 hardware and store them in the \var{READ} array.
    \item  Store in the cyclic buffer: \\
    Copy the contents of \var{READ} into the cyclic Buffer at the current position \var{i},\\
    \var{For k from 0 to Nc - 1:}\\ 
    \hspace*{1.8em}\var{Buffer[(i + k) \% Nb] = READ[k]}
    \item  Retrieve Delayed Data:  \\
    Read \var{Nc} samples from \var{Buffer} into \var{WRITE}, starting from a position shifted by the \var{DELAY} samples relative to the current position \var{i},\\
    \var{For k from 0 to Nc - 1:}\\ 
    \hspace*{1.8em}\var{WRITE [k] = Buffer [(i + k - DELAY) \% Nb]}
    \item  Apply Gain \\
    \var{WRITE = WRITE * Gext}
    \item  Output Signal: \\
    Send the processed voltage samples in \var{WRITE} to the NI 9263 hardware.
    \item  Update Index: \\
    Advance the buffer index pointer for the next loop,\\
    \var{i=i+Nc}
\end{itemize}

\subsubsection*{Theory of $J$-oscillators}

\paragraph{The master equation.} The dynamics of the $J$-oscillator signal can be simulated using a master equation:
\begin{equation}
    \frac{d}{dt} \hat{\rho}(t) = -i [\hat{H}_0 + \hat{V}(t), \hat{\rho}(t)] + \doublehat{R}\hat{\rho}(t) + \hat{P},
\end{equation}
where $\hat{\rho}(t)$ is the density operator, $\hat{H}_0$ represents the $J$-coupling interaction, $\hat{V}(t)$ represent the coupling of the molecule with the feedback loop, $\doublehat{R}$ is the relaxation superoperator, $\hat{P}$ accounts for the SABRE hyperpolarization pumping; $i$ is a unit imaginary number.

\paragraph{Hamiltonian:} The $J$-coupling interaction Hamiltonian for the [$^{15}$N]-ACN maser system is:
\begin{equation}
    \hat{H}_0 = 2 \pi J ( \mathbf{\hat{S}} \cdot \mathbf{\hat{K}} ) ,
\end{equation}
where $\mathbf{\hat{S}}$ and $\mathbf{\hat{K}}$ are the spin operators for $^{15}\mathrm{N}$ and total $^{1}\mathrm{H}$ nuclei, respectively, and $J$ is the scalar coupling constant. The interaction with the feedback magnetic field is modeled as:
\begin{equation}
    \hat{V}(t) =  - G_{\mathrm{ext}}\cdot B_{\mathrm{OPM}}(t-\tau) \cdot (\gamma_{^{15}\mathrm{N}}  \hat{S}_{\mathrm{y}} + \gamma_{^{1}\mathrm{H}}  \hat{K}_{\mathrm{y}}),
\end{equation}
where $G_{\mathrm{ext}}$ is the (external) feedback gain, $\tau$ is externally applied the feedback delay, and $\gamma_{15\mathrm{N}}$, $\gamma_{1\mathrm{H}}$ are gyromagnetic ratios of $^{15}$N and $^{1}$H nuclei, respectively. Consequently, we define $B_{\mathrm{ext}} = G_{\mathrm{ext}}\cdot B_{\mathrm{OPM}}$. The field from the spherically shaped sample as measured by the OPM, $B_{\mathrm{OPM}}(t)$ can be derived from the sample magnetization $M(t)$: 
\begin{equation}
    B_{\mathrm{OPM}}(t) = -\frac{\mu_0}{3} \frac{r^3}{d^3} M(t).
\end{equation}
Here, $d=12.45$\,mm is the distance between the center of the sensor cell and the sample, and $r=4.2$\,mm is the radius of the sample. The magnetization of the system $M(t)$ (would be along the $y$-axis due to the uniaxial nature of zero-field scalar maser) is calculated as,
\begin{equation}
    M(t) = \langle  \gamma_{^{15}\mathrm{N}}  \hat{S}_{\mathrm{y}} + \gamma_{^{1}\mathrm{H}}  \hat{K}_{\mathrm{y}} \rangle (t) \cdot \hbar N_\mathrm{A}  \cdot C   ,
\end{equation}
where $C = 967$\,mM is the [$^{15}$N]-ACN concentration, and $\langle \cdot
\rangle = \mathrm{Tr(\hat{\rho}(t) \cdot)}$ denote the quantum expectation value.

\paragraph{Relaxation.} The relaxation superoperator incorporates intramolecular dipolar interactions, intermolecular interactions, and paramagnetic effects. Intramolecular relaxation is modeled using a rotational diffusion approach for a symmetric top, with tunneling and spinning diffusion times of 0.135\,ps and 3\,ps, respectively. Details of the calculation method will be given elsewhere. 

Intermolecular effects and paramagnetic contributions (e.g., dissolved oxygen) are treated via a random fluctuating field model \cite{levitt1996demagnetization}:
\begin{equation}
    \hat{H}_{\mathrm{RF}}(t)=  -\sum_{j=x,y,z} \left( \gamma_{^{15}\mathrm{N}}\,B_{N \rm j}(t)\,\hat{S}_{\mathrm{j}} +\gamma_{^{1}\mathrm{H}}\,B_{K \rm j}(t)\,\hat{K}_{\mathrm{j}}
\right)\,,
\end{equation}
where $B_{Nj}(t)$ and $B_{Kj}(t)$ are random fields along orthogonal axes. Their correlations satisfy:
\begin{equation}
    \frac{1}{2} \int_{-\infty}^{\infty} dt^{'} \gamma_{^{1}\mathrm{H}}^2 \overline{B_{Ij}(t)B_{I^{'}{j^{'}}}(t-t^{'})} = \frac{1}{T_s} \delta_{jj^{'}} C_{II^{'}}\,,
\end{equation}
with $C_{II^{'}}=1$ for $I=I^{'}$ and $2/3$ otherwise based on the assumption that the noise field applied on $^{15}$N and $^{1}$H spins are partially correlated. The calculation of the corresponding relaxation superoperator is based on the work \cite{xu2023essential}.

Chemical exchange effects from SABRE were found negligible, as evidenced by minimal linewidth differences in 
$J$-spectra with different catalyst concentrations as shown in Fig.\,\ref{fig:FWHMvsPTC}.

\paragraph{SABRE-pumping.} The pumping term $\hat{P}$ ensures steady-state state aligns with the SABRE-hyperpolarized population imbalances at zero-field,
\begin{equation}
    \doublehat{R} \hat{\rho}_{\mathrm{eq}} + \hat{P} = 0\,.
\end{equation}
In the absence of external field, the steady state is isotropic, i.e., it has no preferred directions in space. As a result, the quantum states with the same the total angular momentum ($F$) and total proton angular momentum ($K$) have the same populations. And the steady-state $\hat{\rho}_{\mathrm{eq}}$ can be fully describe by two parameters (see Fig.\,\ref{fig:rhoeq}), $\alpha$ (for population imbalances in $K=3/2$ manifold) and $\beta$ (for population imbalance in $K=1/2$ manifold). The values of are determined from the integrated signals of 1-$J$ and 2-$J$ peaks from a ``ZF-ZF'' experiment \cite{xu2025zero} with a 90$^\circ$ $^{15}$N-$^{1}$H DC pulse: 
\begin{equation}
    \alpha = \frac{I_{\mathrm{1J}}}{8b_0} \,\,\, \beta = \frac{I_{\mathrm{2J}}}{20b_0}
\end{equation}
with,
\begin{equation}
      b_0 = \frac{1}{2} ( \mu_{^{1}\mathrm{H}} 
 -\mu_{^{15}\mathrm{N}}) \cdot C \cdot N_\mathrm{A} \cdot \frac{\mu_0}{3}  \cdot \frac{r^3}{d^3}\,.
\end{equation}
Here, $\mu_{^{1}\mathrm{H}}$ and $\mu_{^{15}\mathrm{N}}$ are the magnetic dipole moment for $^{15}$N and $^{1}$H nucleus, respectively.

\paragraph{The numerical solver.} The master equation is solved via Strang splitting \cite{strang1968construction}, which separates coherent and dissipative dynamics with second-order accuracy. The density operator $\hat{\rho}(t)$ evolves over a timestep $\Delta t$
as follows:
\begin{enumerate}
    \item Coherent half-step: Apply unitary evolution with:
    \[\hat{U}_1 = e^{-i\Bigl(\hat{H}_0 + \hat{H}_{\mathrm{F}}(t+\frac{\Delta t}{4})\Bigr)\frac{\Delta t}{2}},\]
    updating $\hat{\rho} \rightarrow \hat{U}_1  \hat{\rho}  \hat{U}_1^\dagger$.
    
    \item Dissipative step: Update $\hat{\rho}$ via  a Euler step, $\hat{\rho} \rightarrow  \hat{\rho} + \Bigl(\doublehat{R}\hat{\rho} + \hat{P}\Bigr)\Delta t$.

    \item Second coherent half-step: Apply
    \[\hat{U}_2 = e^{-i\Bigl(\hat{H}_0 + \hat{H}_{\mathrm{F}}(t+\frac{3\Delta t}{4})\Bigr)\frac{\Delta t}{2}},\]
    updating $\hat{\rho} \rightarrow \hat{U}_2 \hat{\rho}  \hat{U}_2^\dagger$.
\end{enumerate}

\paragraph{Simulation Parameters:} The model includes three key parameters: $I_{\mathrm{1J}}$, $I_{\mathrm{2J}}$ and $T_{s}$. For Fig.\,\ref{fig:delay and gain}C-\ref{fig:delay and gain}D, we used $I_{\mathrm{1J}}=58$\,pT, $I_{\mathrm{2J}}=106$\,pT and $T_s=28$s. For Fig.\,\ref{fig:delay and gain}E-\ref{fig:delay and gain}F, these parameters were $I_{\mathrm{1J}}=66$\,pT, $I_{\mathrm{2J}}=124$\,pT and $T_s=32$\,s. Note the difference between the two parameter sets arise because the samples were prepared on different days.

\subsubsection*{The intrinsic magnetic gain of samples}
To calculate the intrinsic magnetic gain of the system, $G_{\mathrm{int}}$, we update the master equation to include the interaction between the sample and a weak magnetic field applied via the piercing solenoid. Consider the applied field with magnetic field intensity $H(t)$ that oscillates at frequency $f$. The updated master equation for the system is
\begin{equation}
    \frac{d}{dt} \hat{\rho}(t) = -i [\hat{H}_0 + \hat{V}(t), \hat{\rho}(t)] + \doublehat{R}\hat{\rho}(t) + \hat{P},
\end{equation}
with
\begin{equation}
    \hat{V}(t) = - \mu_0 H(t) \cdot (\gamma_{^{15}\mathrm{N}}  \hat{S}_{\mathrm{y}} + \gamma_{^{1}\mathrm{H}}  \hat{K}_{\mathrm{y}}).
\end{equation}
At the dynamic steady state, the magnetization $M(t)$ of the sample oscillates at the same frequency $f$ as the applied field $H(t)$. This magnetization generates a detectable magnetic flux at the OPM. When the applied field amplitude is sufficiently small (i.e.,  when the perturbation from the applied field is weak compared to the system’s relaxation rate), the ratio between the amplitudes of the measured magnetic flux (at the sensor) and the applied magnetic flux becomes independent of the applied field amplitude. Instead, it depends solely on the frequency $f$. 

We define the internal magnetic amplification $G_{\mathrm{int}}(f)$ as this frequency-dependent ratio, expressed mathematically as
\begin{equation}
    G_{\mathrm{int}}(f) =  \frac{r^3}{3 d^3} \cdot \frac{| M(t) |}{| H(t)|} = \frac{r^3}{3 d^3} | \chi (f) | ,
\end{equation}
where $r$ is the radius of the sample and $d$ is the distance between the sensor and sample, $| \chi (f) | $ is a frequency-dependent magnetic susceptibility.

Simulation shows maximal internal gains at $f=J$ and $f=2J$, with internal gains $G_{\mathrm{int}}(J)\approx13.8\,\%$ and  $G_{\mathrm{int}}(2J)\approx17.2\,\%$, respectively (Fig.\,\ref{fig:Gin}). 

As discussed, the threshold external feedback gain ($G^{\mathrm{th}}_{\mathrm{ext}}$) required for maser oscillations to spontaneously emerge corresponds to the inverse of the on-resonance internal gains:
\begin{equation}
    G^{\mathrm{th}}_{\mathrm{ext}} = \frac{1}{G_{\mathrm{int}}(f_0)}.
\end{equation}
For 1-$J$ and 2-$J$ transitions, this yields $G^{\rm th}_{\mathrm{ext}}\approx7.2$ and $G^{\rm th}_{\mathrm{ext}}\approx5.8$, respectively, consistent with the results in Fig.\,\ref{fig:delay and gain}E-\ref{fig:delay and gain}F.

\subsubsection*{The normalization of Fourier Transformation}
Suppose a continuous signal \( x(t) \) is sampled at a rate \( f_s \) to get a discrete sequence
\begin{equation}
    x[n] = x\Bigl(t=\frac{n}{f_s}\Bigr),\quad n = 0,1,\dotsc,N-1,
\end{equation}
over a total duration \( T = N/f_s \). The discrete Fourier transform (DFT) is then calculated as, 
\begin{equation}
    X[k] = \sum_{n=0}^{N-1} x[n]\, e^{-i\frac{2\pi k}{N} n}, \quad k = 0,1,\dotsc,N-1.
\end{equation}
Interpreting the sum as a Riemann approximation, i.e., $\Delta t \cdot \sum (\cdot) \approx \int_{0}^{T} (\cdot) \, dt$, $X[k]$ approximates the continuous Fourier integral,
\begin{equation}
    X[k] \approx f_s \int_{0}^{T} x(t)\, e^{-i2\pi f_k t}\, dt,
\end{equation}
with the discrete frequency \( f_k = \frac{k f_s}{N} \) and the sampling interval \( \Delta t = 1/f_s \).

The two different normalization convention adopted in the main work are:

\paragraph{Convention I:} Divide \( X[k] \) by the number of samples \( N \), giving,
\begin{equation}
    \widetilde{X}_{\mathrm{I}}[k] = \frac{X[k]}{N} \approx \frac{1}{T}\int_{0}^{T} x(t)\, e^{-i2\pi f_k t}\, dt.
\end{equation}
In this case, the Fourier amplitudes retain the original units of \( x(t) \) (e.g., pT). This convention is adopted in Fig.\,\ref{fig:diverse molecules}.

\paragraph{Convention II:} Multiply \( X[k] \) by \( \Delta t = 1/f_s \), so that,
\begin{equation}
    \widetilde{X}_{\mathrm{II}}[k] = \frac{X[k]}{f_s} \approx \int_{0}^{T} x(t)\, e^{-i2\pi f_k t}\, dt.
\end{equation}
This yields Fourier amplitudes with units such as pT/Hz (if \( x(t) \) is measured in pT). This convention was adopted in Fig.\,\ref{fig:schematics}





\newpage

\begin{figure}
	\centering
	\includegraphics[width=15.4cm]{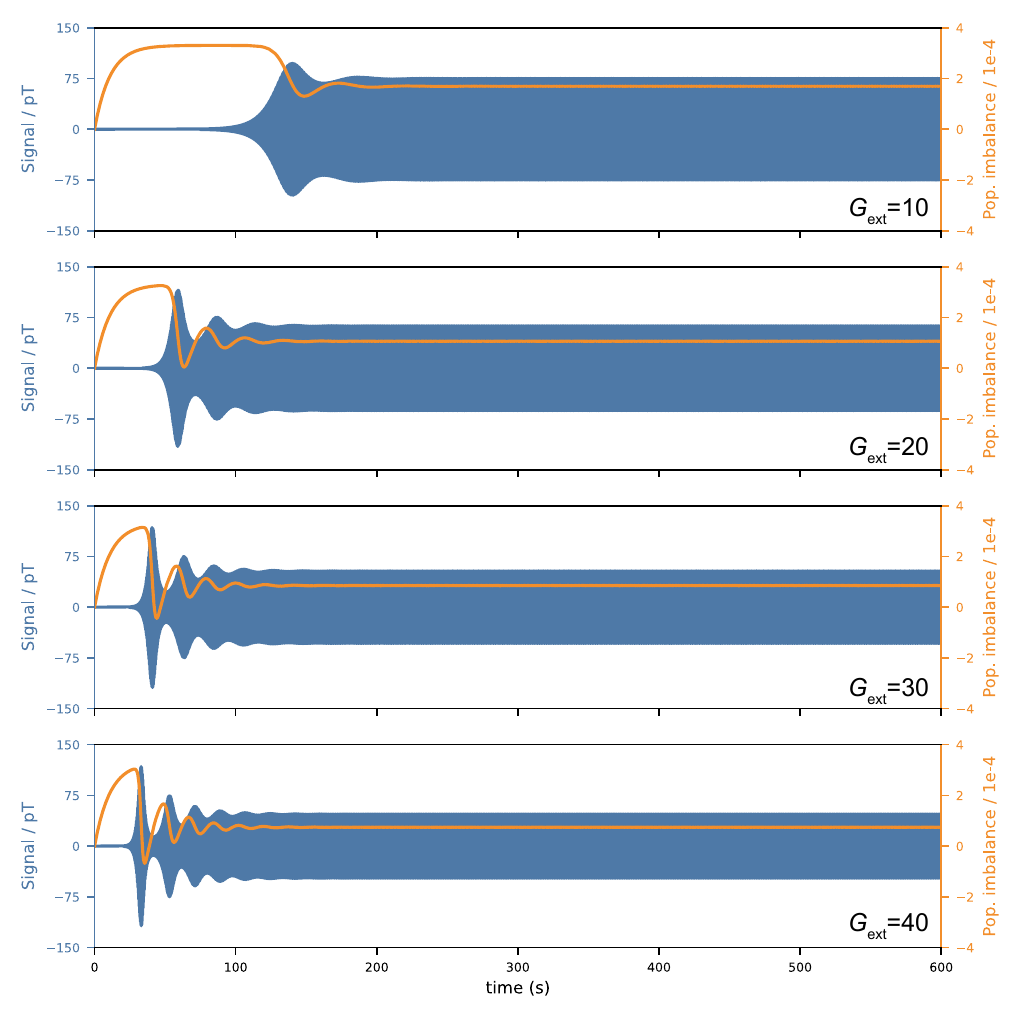} 
	\caption{\textbf{Simulations of $J$-oscillators as a function of the feedback gain.} The blue trace represents the oscillator signal while the orange trace shows the population difference, calculated as (3/4) × (population in $K=3/2,\,F=2$ states) - (5/4) × (population in $K=3/2,\,F=1$ states); the scaling factors (3/4 and 5/4) account for the different number of available states in $F=2$ versus $F=1$ (see Fig.\,\ref{fig:rhoeq}). The feedback delay was fixed to $\tau=222$\,ms; rms of the OPM noise was set to 0.1\,pT. As the feedback gain increases, the initial SABRE-pumped population imbalance can be inverted during the burst events.}
	\label{fig:overshooting}
\end{figure}

\begin{figure}
	\centering
	\includegraphics[width=8.5cm]{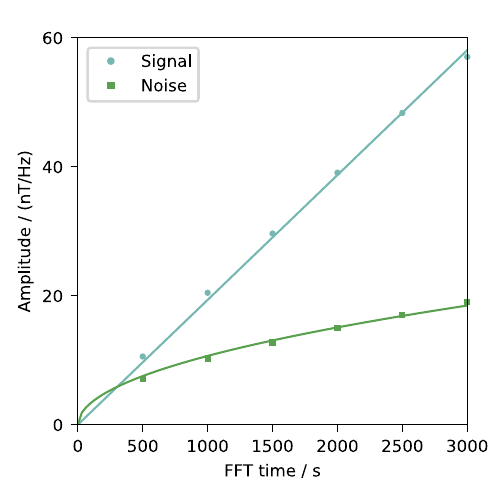} 
	\caption{\textbf{The spectral amplitude and noise floor versus FT duration.} The FT spectra were obtained from the time-domain quantum oscillator signal (Fig.\,1D) in using sliding time windows starting at 100\,s with increasing durations. The noise floors were sampled over 8-10\,Hz range. Solid lines are fits to the data using a linear function for the spectral amplitude and a square-root function for the noise floor. Consequently, the SNR of the oscillator signal scales as the square-root of the acquisition time. }
	\label{fig:scaling} 
\end{figure}

\begin{figure}
	\centering
	\includegraphics[width=8.5cm]{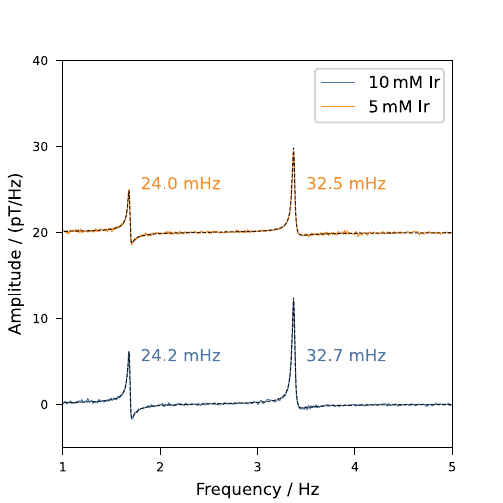} 
	\caption{\textbf{Dependence of the signal linewidths on the catalyst concentrations.} Zero-field spectra of naturally abundant [$^{15}$N]-ACN\,/\,[$^{14}$N]-ACN samples (0.36\,\%\,/\,99.6\,\%) with catalyst concentrations of 5\,mM (orange) and 10\,mM (blue). The co-ligand (benzylamine) to catalyst concentration ratio was 25 for both samples. Dashed lines closely following the experimental data represent dual Lorentzian fits, from which the full-width-at-half-maximum of the corresponding peaks was extracted. }
	\label{fig:FWHMvsPTC}
\end{figure}

\begin{figure}
	\centering
	\includegraphics[width=8.5cm]{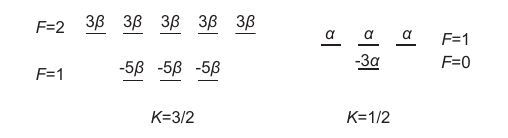} 
	\caption{\textbf{Visualization of $\hat{\rho}_{\mathrm{eq}}$ at zero-field.} The parameters $\alpha$ and $\beta$ (with coefficients) denote the offsets of populations with respect to the thermal state at zero-field. The states with the same total proton angular momentum $K$ and total angular momentum $F$ have identical populations.}
	\label{fig:rhoeq}
\end{figure}

\begin{figure}
	\centering
	\includegraphics[width=8.5cm]{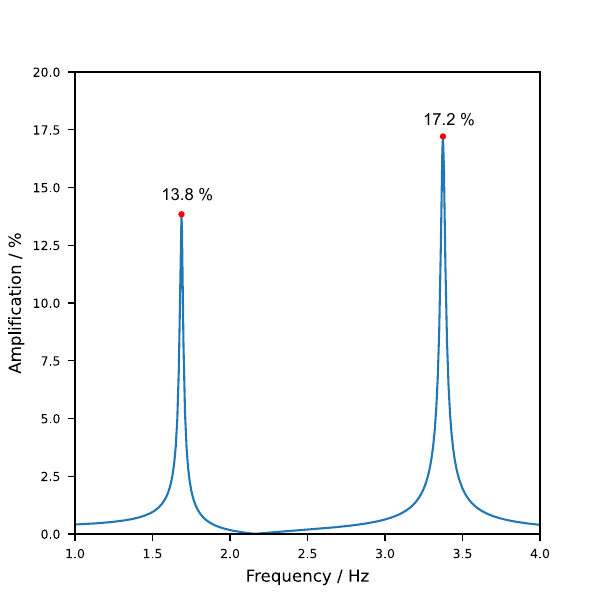} 
	\caption{\textbf{Simulated dependence of the intrinsic gain ($G_{\mathrm{int}}$) of the system on the frequency ($f$) of AC excitation.}  The simulations show maximal amplification at $f=J$ and $f=2J$ ($J=1.687$\,Hz), with internal gains $G_{\mathrm{int}}(J)\approx13.8\,\%$ and  $G_{\mathrm{int}}(2J)\approx17.2\,\%$, respectively. The same parameters were adopted here as used for the simulation in Fig.\,\ref{fig:delay and gain}E-\ref{fig:delay and gain}F.}
	\label{fig:Gin}
\end{figure}

\begin{figure}
	\centering
	\includegraphics[width=15.4cm]{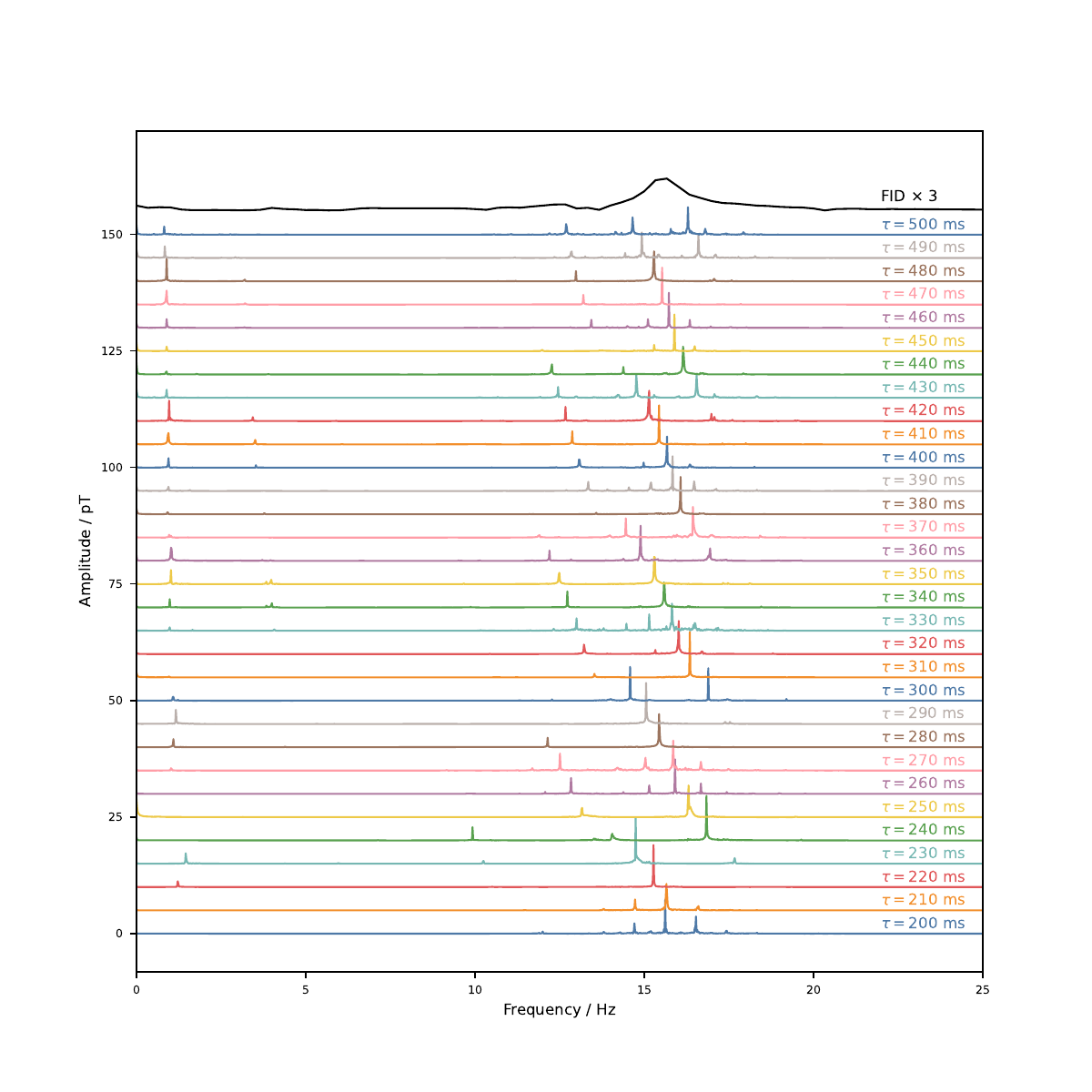} 
\caption{\textbf{Stacked spectra from quantum oscillators in [$^{15}$N]-pyridine acquired at varying feedback delays ($\tau$).} The external feedback gain was fixed at $G_{\mathrm{ext}}=-3000$ for all experiments. Each spectrum corresponds to a 1~min acquisition, with Fourier transformation applied to the time-domain data from 5–60~s to generate the stacked spectra. The top spectrum shows a conventional zero-field NMR spectrum of the same sample, for reference. The observed oscillations are discussed in the main text and match the predictions from the simplified numerical simulations shown in Fig.\,\ref{fig:2spin_simu}.}
	\label{fig:pyridine_oscillator}
\end{figure}

\begin{figure}
	\centering
	\includegraphics[width=15.4cm]{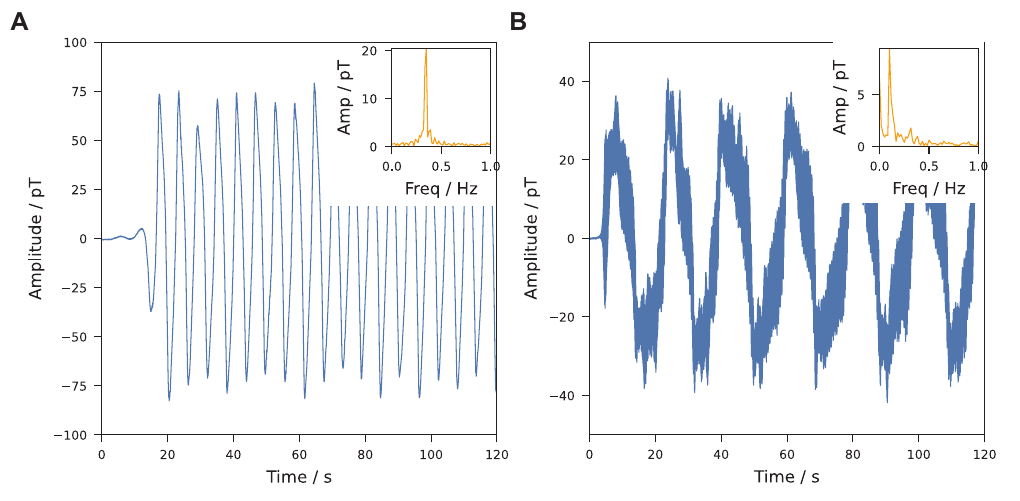} 
	\caption{\textbf{The $J$-oscillators operating on  heterocycle molecules generate ultralow frequency signals.} (A) [$^{15}$N$_3$]-metronidazole oscillator acquired with $G_{\mathrm{ext}}=2000$ and $\tau=140$\,ms; (B) 4-amino[$^{15}$N]-pyridine oscillator acquired with $G_{\mathrm{ext}}=1500$ and $\tau=130$\,ms. Insets in both panels show the Fourier Transform of the oscillator signal over the 20–120\,s time window. }
	\label{fig:heterocyclic_ultralow}
\end{figure}

\begin{figure}
	\centering
	\includegraphics[width=15.4cm]{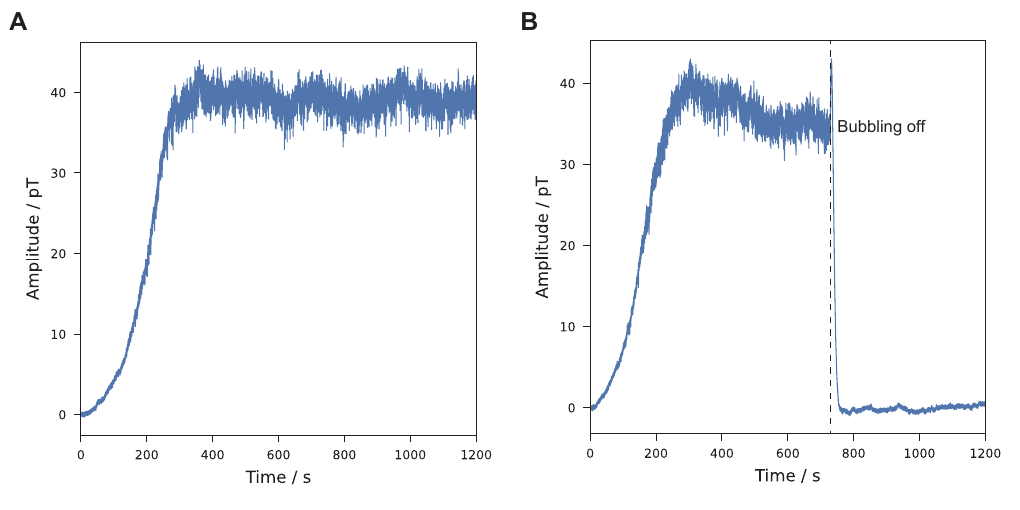} 
	\caption{\textbf{The $J$-oscillator on [U-$^{13}$C, $^{15}$N]-butyronitrile.} The data were acquired with $G_{\mathrm{ext}}=-2000$ and $\tau=150$\,ms. (A) A 20-min acquisition produces a DC maser signal (no oscillations are resolvable within the acquisition window). (B) The signal acquired under the same conditions, but with \textit{para}-H$_2$ bubbling stopped at around 700\,s (marked as the dashed line). The signal disappears upon stopping the bubbling, confirming that its origin relates to sample hyperpolarization and feedback.}
	\label{fig:chain_DC}
\end{figure}

\begin{figure}
	\centering
	\includegraphics[width=15.4cm]{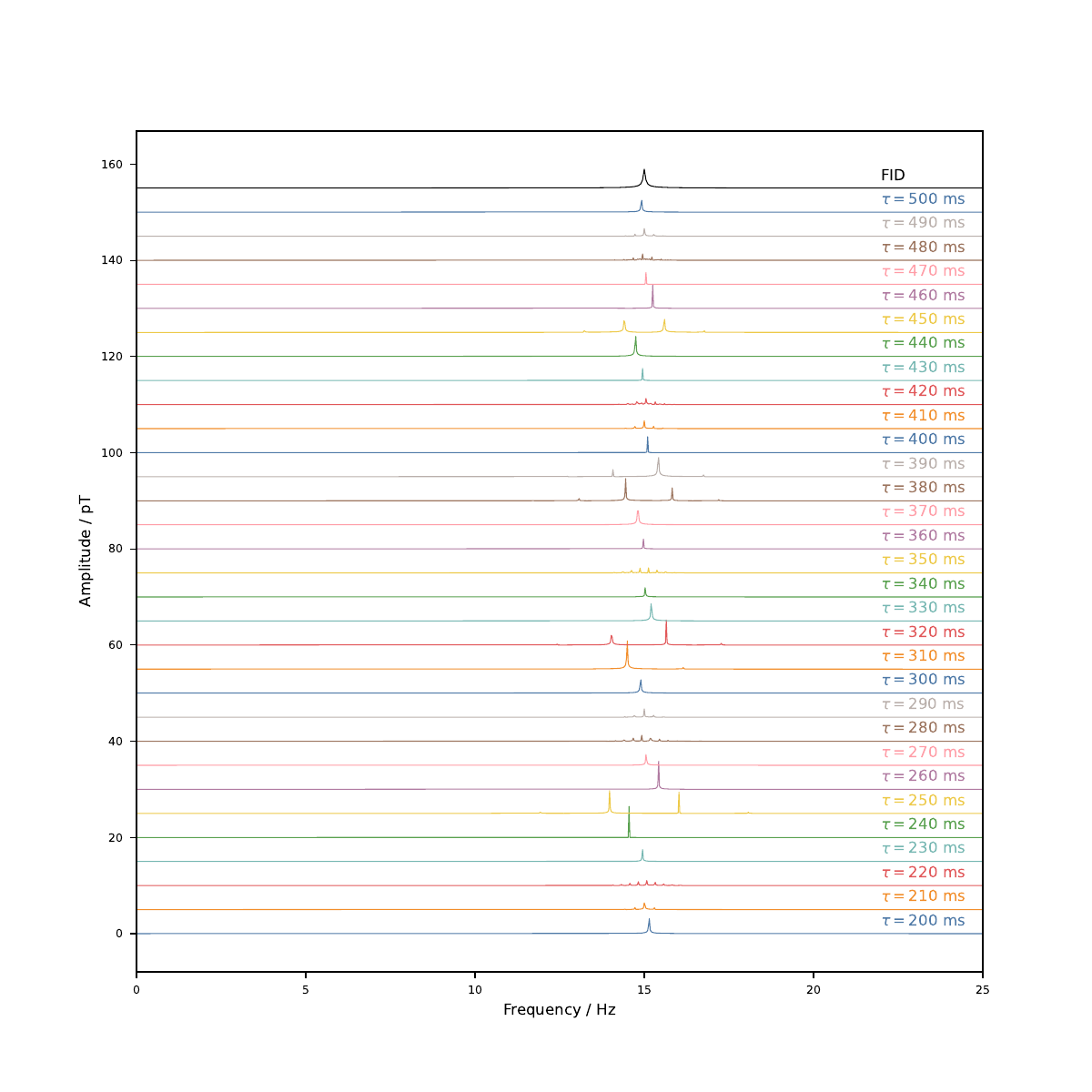} 
	\caption{\textbf{Simulations of $J$-oscillators operating on coupled $^{15}$N-$^{1}$H two spin system.} Simulations assume a scalar $J$ coupling constant of 15\,Hz, OPM rms noise of 0.1\,pT and identical system geometry to previous experiments. The SABRE-pumped population imbalances are set such that the integral of peak in the conventional zero-field NMR measurement equals to 50\,pT. The nuclear spin relaxation are accounted using random fluctuating field model, resulting in resonance linewidths (FWHM) of 0.2\,Hz in conventional $J$-spectra. The external feedback gain is fixed at $G_{\mathrm{ext}}=-3000$ for all simulation. Each spectrum corresponds to a 1\,min acquisition, with Fourier transformation applied to the time-domain data from 20–60\,s to generate the stacked spectra. The top spectrum shows a simulation of conventional zero-field NMR spectrum of the system, for reference. }
	\label{fig:2spin_simu}
\end{figure}

\begin{figure}
	\centering
	\includegraphics[width=8.4cm]{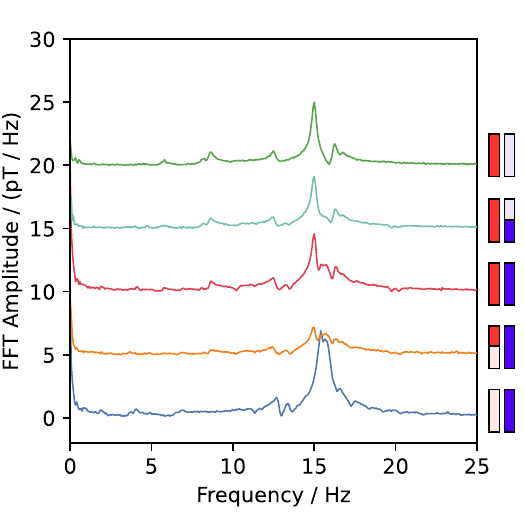} 
	\caption{\textbf{Zero-field NMR spectrum of five pyridine / 4-aminopyridine mixtures tested in Fig.\,\ref{fig:mixture}}. The compositions of each sample are denoted by the same defined bar pairs as in Fig.\,\ref{fig:mixture}, which indicate the specific $^{15}$N isotopic enrichment of both pyridine and 4-aminopyridine substrates.}
	\label{fig:mix_fid}
\end{figure}

\begin{figure}
	\centering
	\includegraphics[width=15.4cm]{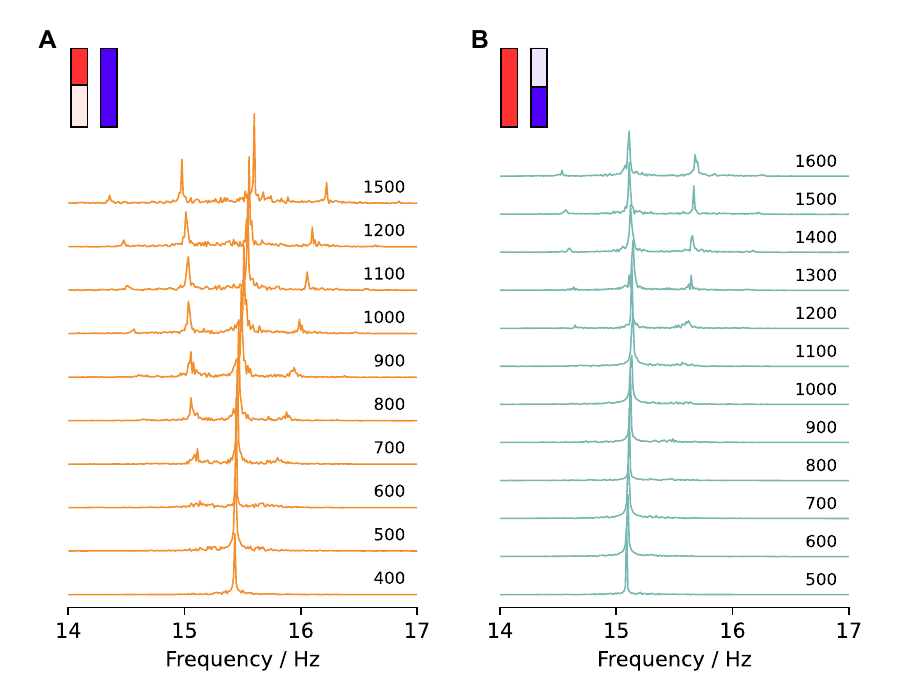} 
	\caption{\textbf{Quantum $J$-oscillators operated on pyridine / 4-aminopyridine mixture with different $^{14}$N/$^{15}$N isotopic compositions.} The samples are denoted by the same defined bar pairs as in Fig.\,\ref{fig:mixture}. (A)~ Sample with 100~\% $^{15}$N-enriched pyridine and 50~\% $^{15}$N-enriched 4-aminopyridine. (B)~Sample with 50~\% $^{15}$N-enriched pyridine and 100~\% $^{15}$N-enriched 4-aminopyridine. The feedback delay is fixed at 115\,ms for all measurements and the applied feedback gain is annotated to the left of each trace.  }
	\label{fig:mix_SI}
\end{figure}

\newpage
\newpage
\begin{table}
\centering
\caption{\textbf{External feedback phase lags at oscillator frequencies for given delay intervals.} The delay intervals show where the quantum oscillator on [$^{15}$N]-ACN is sustained, as extracted from Fig.\,\ref{fig:delay and gain}. Parentheses indicate the initial or end sampling intervals rather than real cutoffs. The phase lag is derived using $\varphi = 2\pi f \tau$ at around the $J$-transition frequencies $f\!=\!J$ or $f\!=\!2J$ ($J\!=\!1.687$~Hz), accounting for frequency-dependent phase lags due to delay. For the negative external feedback gain ($G_{\text{ext}} = -20$ ), an additional $\pi$ phase lag is added.\\}
\label{tab:phase_lags}

\begin{tabular}{lcccc}
\hline
\textbf{$G_{\text{ext}}$} & \textbf{Oscillator} & \textbf{Delay (ms)}  & \textbf{Phase Lag $\varphi$ (rad)} \\ 
\hline
-20 & 1-$J$ & $\langle 60 \rangle$--275  & $\langle 3.78 \rangle$--6.05 \\ 
    & 2-$J$ & $\langle 60 \rangle$--140  & $\langle 4.41 \rangle$--6.11 \\ 
\hline
~20 & 1-$J$ & 325--$\langle 400 \rangle$  & 3.44--$ \langle 4.24 \rangle$ \\ 
    & 2-$J$ & 160--290  & 3.39--6.14 \\ 
\hline
\end{tabular}
\end{table}

\begin{table} 
    \centering
    \caption{\textbf{Sample compositions for $J$-oscillators; model (Figs.\,1-2) and molecules A-I (Fig.\,3); PTC = polarization transfer catalyst.}}
    \label{tab:sample_compositions}
    \begin{tabular}{l l c l l} 
        \hline
        \textbf{Sample} & \textbf{Substrate} & \textbf{PTC [mM]} & \textbf{Co-substrate} & \textbf{Solvent} \\
        \hline
        Model & [$^{15}$N]-ACN (5\,\%) & 5.0 & BnNH$_2$ (125.0\,mM) & ACN \\
        A & [$^{15}$N]-ACN (0.36\,\%) & 5.0 & BnNH$_2$ (125.0\,mM) & ACN \\
        B & [1-$^{13}$C,$^{15}$N]-ACN (1.0\,\%) & 5.3 & BnNH$_2$ (125.0\,mM) & ACN \\
        C & [2-$^{13}$C,$^{15}$N]-ACN (1.0\,\%) & 5.1 & BnNH$_2$ (125.0\,mM) & ACN \\
        D & [$^{15}$N]-pyridine (100.0\,mM) & 5.0 & - & MeOH \\
        E & [$^{15}$N$_2$]-imidazole (97.1\,mM) & 4.9 & - & MeOH \\
        F & [$^{15}$N$_3$]-metronidazole (53.5\,mM) & 4.2 & - & MeOH-d\textsubscript{4} \\
        G & 4-amino[$^{15}$N]-pyridine (92.9\,mM) & 4.4 & - & MeOH \\
        H & [1-$^{13}$C]-pyruvate (76.8\,mM) & 6.3 & DMSO (24.1\,mM) & MeOH \\
        I & [U-$^{13}$C, $^{15}$N]-butyronitrile (107.4\,mM) & 5.5 & pyridine (52.4\,mM) & MeOH/MeOH-d\textsubscript{4} (1:1) \\
        \hline
    \end{tabular}
\end{table}


\clearpage 





\end{document}